\documentclass{aa}  

\usepackage{natbib}
\bibpunct{(}{)}{;}{a}{}{,}

\usepackage{graphicx}

\usepackage{txfonts}

\usepackage[]{hyperref}

\usepackage{color}
\usepackage{orcidlink}
\begin{document}

   \title{Phase-Induced Amplitude Apodization Complex Mask Coronagraph (PIAACMC) on-sky demonstration with MagAO-X\thanks{This paper includes data gathered with the 6.5 metre Magellan Telescopes located at Las Campanas Observatory, Chile.}}

   \subtitle{}

   \author{E. Tonucci
          \inst{1}\,\orcidlink{0009-0006-4370-822X}
          \and
          S. Y. Haffert\inst{1}\fnmsep\inst{2}\,\orcidlink{0000-0001-5130-9153}
          \and
          W. B. Foster\inst{2}
          \and
          J. R. Males\inst{2}\,\orcidlink{0000-0002-2346-3441}
          \and
          O. Guyon\inst{2,3,4,5}\,\orcidlink{0000-0002-1097-9908}
          \and
          L. M. Close\inst{2}\,\orcidlink{0000-0002-2167-8246}
          \and
           K. Van Gorkom\inst{2}\,\orcidlink{0000-0002-4877-7762}
          \and
           A. D. Hedglen\inst{6}
           \and
           P. T. Johnson\inst{2}
           \and
           M. Y. Kautz\inst{2}\,\orcidlink{0000-0003-3253-2952}
           \and
           J. K. Kueny\inst{3}\,\orcidlink{0000-0001-8531-038X}
           \and
           J. Li\inst{2}\,\orcidlink{0000-0002-8110-7226}
           \and
            J. Liberman\inst{3}\,\orcidlink{0000-0002-4934-3042}
           \and
            J. D. Long\inst{7}\,\orcidlink{0000-0003-1905-9443}
           \and 
           J. Lumbres \inst{3}
           \and
            M. Mars \inst{1}\,\orcidlink{0000-0001-7353-3391}
           \and
           E. A. McEwen \inst{3}\,\orcidlink{0000-0003-0843-5140}
           \and
           A. McLeod \inst{8}
           \and
           L. A. Pearce \inst{9}\,\orcidlink{0000-0003-3904-7378}
           \and
           L. Schatz\inst{10}
           \and
           K. Twitchell\inst{3}\,\orcidlink{0009-0002-9752-2114}
          }

   \institute{Leiden Observatory, Leiden University, PO Box 9513, 2300 RA, Leiden, The Netherlands\\ \email{tonucci@strw.leidenuniv.nl}
         \and
             {Steward Observatory, The University of Arizona, 933 North Cherry Avenue, Tucson, Arizona}
        \and
        Wyant College of Optical Sciences, The University of Arizona, 1630 E University Blvd, Tucson, Arizona
        \and
        Subaru Telescope, National Observatory of Japan, National Institutes of Natural Sciences, 650 N. A'ohoku Place, Hilo, Hawai'i
        \and
        Astrobiology Center, National Institutes of Natural Sciences, 2-21-1 Osawa, Mitaka, Tokyo, Japan
        \and
        Northrop Grumman Corporation, 600 South Hicks Road, Rolling Meadows, Illinois
        \and
        Center for Computational Astrophysics, Flatiron Institute, 162 5th Avenue, New York, New York
        \and
        Draper Laboratory, 555 Technology Square, Cambridge, Massachusetts
        \and
        Department of Astronomy, University of Michigan, Ann Arbor, MI 48109, USA
        \and
         Starfire Optical Range, Kirtland Air Force Base, Albuquerque, New Mexico         
    }

   \date{Received 28 October 2025 / Accepted 20 February 2026}

  \abstract 
   {Advancing the technological development of small inner working angle (IWA) coronagraphs is essential to enabling high-contrast imaging of temperate exoplanets with future extremely large telescopes (ELTs). The PIAACMC has been shown to closely approach the theoretical limit for coronagraphic throughput but its performance has not been fully characterised on-sky.
   This study serves as the first on-sky characterisation of contrast and IWA performance of the PIAACMC. We also performed a technological demonstration of the PIAACMC at sub-micron nearinfrared (NIR) wavelengths approaching the visible.
   We designed and manufactured phase-shifting focal plane masks for the PIAACMC. These were optimised for two cases, a narrowband 875 filter (875 nm, $\sim$3\% bandwidth) and a broadband z' filter (908 nm, $\sim$14\% bandwidth). We tested the coronagraphs both with an internal source and on-sky using MagAO-X, the extreme adaptive optics (XAO) instrument for the Magellan Clay 6.5 m telescope at Las Campanas Observatory, Chile.
   We show good recovery of the off-axis light's point spread function (PSF) shape both with the internal source (within $\sim$92\% and $\sim$97\% depending on the separation) and on-sky when aligning the inverse set of PIAA lenses. We demonstrate sub-$\lambda$/D IWAs with the instrument's internal source of about 0.74 $\lambda$/D with the 875 filter and 0.76 $\lambda$/D with the z' filter. We reach average raw contrasts within 1 and 5 $\lambda$/D with the internal source of about 1.6$\times$10\textsuperscript{$-$3} with the 875 filter and 1.3$\times$10\textsuperscript{$-$3} with the z' filter. These are mainly limited by the focal plane mask manufacturing errors, jitter, and residual quasi-static speckles in MagAO-X. We also show on-sky average raw contrasts within 1 and 5 $\lambda$/D of about 1.4$\times$10\textsuperscript{$-$2} with the 875 filter and 7.8$\times$10\textsuperscript{$-$3} with the z' filter. These are likely limited by wavefront control, low-order aberrations, and poor observing conditions.
   We have successfully characterised PIAACMC's performance and demonstrated its technology on-sky for the first time at sub-micron wavelengths. Future work will improve the design and manufacturing processes of the focal plane masks to improve robustness and reach deeper contrast, as well as integrate focal plane wavefront control for non-common path aberrations (NCPAs) correction.}

   \keywords{Instrumentation: adaptive optics -- Instrumentation: high angular resolution -- Techniques: high angular resolution}
   
   \titlerunning{PIAACMC on-sky demonstration with MagAO-X}
   \maketitle

\section{Introduction}
High-contrast imaging (HCI) is an observation technique that allows us to spatially resolve the close-in circumstellar environment (for a general introduction on HCI and its data reduction strategies, see \citet{follette2023}).
With HCI, we have currently observed about 80 sub-stellar companions. These are mostly young (of the order of 1 to 10 Myr), self-luminous, and have large orbital separations of the order of 10 to 100 AU (for a review on planets detected through direct imaging, see \citet{chauvin2023,zurlo2024}). The only exceptions are the colder gas giant exoplanets Epsilon Indi Ab and 14 Herculis, recently observed by the James Webb Space Telescope (\citet{EpsIndiA},\citet{14Herc}). HCI is an effective technique to study the dynamics of planetary systems \citep{wang2018,rickman2022} and better understand planet formation \citep{haffert_PDS70,richelle_letter2025,close_letter2025}. However, the most important scientific goal within the exoplanet field is arguably characterising atmospheres of smaller, older planets, especially those in the habitable zones of their host star that could possibly host life. This closely relates to the themes of the European Space Agency (ESA) Voyage 2025 (see \citet{ESAvoyage}) and the National Academies of Sciences, Engineering, and Medicine's 2020 Decadal Survey \citep{decadal_survey}. Moreover, it aligns with the driving science cases of the future extremely large telescopes (ELTs), with HCI instruments such as the Planetary Camera and Spectrograph (PCS, \citet{PCS}) for the European Extremely Large Telescope (E-ELT, \citet{ELT}), and the Giant Magellan Adaptive Optics eXtreme (GMagAO-X, \citet{GMagAO-X}) instrument for the Giant Magellan Telescope (GMT, \citet{GMT}). Characterising atmospheres of temperate exoplanets is therefore a key driver for the technological development of telescopes and instruments of the future. HCI is a particularly promising technique to characterise atmospheres of such planets, thanks to the spatial separation between star and planet \citep{snellen2015,follette2023}. Indirect techniques (e.g. transit spectroscopy), on the other hand, are limited by the challenge of disentangling planet and stellar signals and modelling stellar noise and variability \citep{disentangle_star}.

As the name suggests, the first challenge of HCI is contrast, the ratio between the flux of the companion and the flux of its host star. For an exoplanet around a Sun-like star, contrast can reach values of 10\textsuperscript{$-$6} to 10\textsuperscript{$-$10} for a Jupiter-like and Earth-like planet, respectively \citep{matt_sebastiaan_HCC}. Coronagraphs are used to observe the planet light that would otherwise be hidden under the glare of the star. A coronagraph is an optical system acting as an angular filter; it suppresses the on-axis stellar light and lets the off-axis planet light through. Many coronagraph types and concepts have been proposed in the last decades (for a comprehensive review of high-contrast coronagraphy, see \citet{matt_sebastiaan_HCC}). Keeping in mind the interest in characterising Earth-like planets in the habitable zone, coronagraphs worth pursuing are those with high throughput and small IWA. The IWA is defined as the smallest separation where the throughput is 50\% of the maximum reachable throughput of the system.

\citet{Guyon2006} compared simulated performances of different coronagraphs for both a point source and an extended source with an angular size still smaller than the diffraction limit, and found an upper limit in throughput imposed by physics. The coronagraph type that most closely approaches the theoretical limit is the Phase-Induced Amplitude Apodization Coronagraph (PIAAC, \citet{Guyon2005}). The Phase-Induced Amplitude Apodization (PIAA) technique \citep{Guyon2003} is performed via two sets of optics (mirrors or lenses) placed at the entrance and at the exit of the coronagraphic system that use pupil reshaping to produce an apodised pupil. The best performing coronagraph of this type is the PIAACMC \citep{Guyon2010}, which uses lossless aspheric lenses and a phase-shifting focal plane mask. The PIAACMC theoretically offers nearly 100\% throughput and can reach sub-$\lambda$/D IWA, making it a coronagraph that can operate at the diffraction limit.

Another challenge comes from diffractive elements in the telescope aperture such as obscuration, spiders, and gaps between segmented mirrors. These contaminate the science plane with stellar light leakage, causing coronagraphic performance to drop. Segmented mirrors and off-axis telescope designs are predominant in the design of future observatories both on the ground and in space, being an important driver for coronagraph technological development. \citet{Guyon2013} demonstrated that the PIAACMC maintains high simulated performance for any telescope aperture shape, making it one of the most promising coronagraphs to be employed in future observatories.

Since its introduction, different PIAACMCs have been designed, manufactured and tested in the laboratory. At the High Contrast Imaging Testbed (HCIT) of the Jet Propulsion Laboratory in Pasadena, USA, \citet{Marx2021} reached a contrast of 3$\times$10\textsuperscript{$-$8} within 4 to 9 $\lambda$/D at 2\% bandwidth. Later, \citet{Belikov2022} showed a contrast of 1.9$\times$10\textsuperscript{$-$8} within 3.5 to 8 $\lambda$/D at 10\% bandwidth, again at HCIT. They have tested segmented and obscured apertures, and the focal plane mask was fabricated using lithography. PIAACMC testing is also performed with the Segmented pupil Experiment for Exoplanet Detection testbed (SPEED) at the Lagrange Laboratory of Université Côte d’Azur in Nice, France \citep{Martinez2020}. Finally, PIAACMC is also fully implemented at the Subaru Coronagraphic Extreme Adaptive Optics system (SCExAO) \citep{Lozi2009,Lozi2018}. A preliminary demonstration of two PIAACMCs was performed in near infrared, in H band. \citet{lozi_poster} showed a laboratory throughput curve with an estimated IWA of 58 mas, corresponding to 1.43 $\lambda$/D at 1.55 $\mu$m. At the same separation they reached a contrast of $\sim$1.5$\times$10\textsuperscript{$-$3}. Later, an on-sky observation of Eta Virginis A and B, stars in a triple system, was also performed \citep{justin_thesis}. The analysis, however, was limited to estimating the separation between the two stars and their flux ratio.

This paper presents the first on-sky characterisation of contrast and IWA performance of the PIAACMC. This demonstration is performed in sub-micron NIR with the Magellan Adaptive Optics eXtreme instrument (MagAO-X, \citet{magaox_newest}). This is framed in the context of the development of coronagraphy at 1 to 2 $\lambda$/D to perform reflected light science of our nearest neighbour, Proxima Centauri b. We validate the PIAACMC's performance both in `lab mode' with the internal source of the instrument, and on-sky. Demonstrating and advancing the PIAACMC's technology at current facilities is of pivotal importance in view of future ground-based and space-based observatories, including the above mentioned E-ELT and GMT, as well as the Habitable Worlds Observatory (HWO). In section \ref{sec:system}, we introduce the MagAO-X system and the layout of the PIAACMC as proposed in \citet{Guyon2010}. In section \ref{sec:design_manufacturing}, we discuss the design and manufacturing processes of the PIAACMC components for MagAO-X. Section \ref{sec:lab_results} shows the performance results of the laboratory testing. We present the on-sky performance results in Section \ref{sec:onsky_results}.
To conclude, in section \ref{sec:conclusions} we sum up the most important considerations of this study and discuss future prospects.

\section{System and coronagraph}\label{sec:system}

\subsection{MagAO-X}\label{subsec:magaox}
MagAO-X \citep{magaox_newest} is an XAO system optimised for HCI at visible and sub-micron NIR wavelengths for the Magellan Clay 6.5 m telescope at Las Campanas Observatory, Chile. It delivers high Strehl ratios (>70\%) at high resolutions (14-30 mas) and high contrasts (<10\textsuperscript{$-$4} from $\sim$1 to 10 $\lambda$/D). Its main scientific goal is the detection and characterisation of Solar System-like exoplanets. Other objectives include the characterisation of young gas giants, circumstellar discs, and asteroids, with the aim of better understanding early stages of planet formation, as well as the spectroscopy of resolved stellar surfaces. Young planets that are still accreting are detected in H$\alpha$ \citep{close2013,haffert_PDS70,richelle_letter2025,close_letter2025,Li2025}, while older, temperate planets are planned to be characterised in reflected light.

\begin{figure}[h!]
    \centering
    \includegraphics[width=0.98\linewidth]{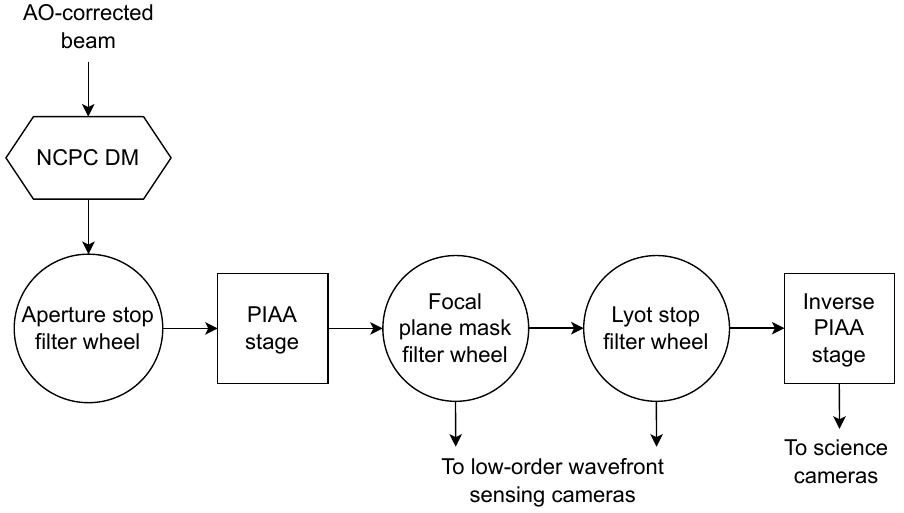}
    \caption{Schematic representation of MagAO-X's coronagraphic arm. The AO-corrected beam is sent to the NCPC deformable mirror; then, it passes through a filter wheel containing different aperture stops; a stage with the forward PIAA lenses set can be moved in or out of the beam through motorised actuators; light is then focused on a filter wheel containing different focal plane masks; another filter wheel follows with different Lyot stops; any light reflected from the focal and pupil planes is sent to separate low-order wavefront sensing cameras; another motorised stage containing the inverse PIAA lenses set can be moved in and out of the beam; finally, light is sent to the science cameras.}
    \label{fig:cor_system}
\end{figure}

The adaptive optics (AO) system consists of a combination of two deformable mirrors (DMs). The first is an Alpao DM97-15, a low-order DM with 97 actuators that serves as the `woofer', correcting for low-order modes and reducing the necessary stroke on the second DM. The second DM is a Boston MicroMachines (BMC) 2K high-order Micro-Electro-Mechanical Systems (MEMS) DM that serves as the `tweeter'. This DM has 2\,040 actuators and corrects for higher order modes \citep{woofer-tweeter}. A physical obstruction left over from the lithography process prevents the surface from fully deflecting causing a `bump' on this DM that must be masked for coronagraphic imaging.
After an atmospheric dispersion corrector (ADC), the light is split between a wavefront sensing arm and a coronagraphic arm. In the wavefront sensing arm, a pyramid wavefront sensor (PWFS) is used for primary wavefront sensing, reaching a correction speed of up to 3.63 kHz. The coronagraphic arm's layout is schematically shown in Fig. \ref{fig:cor_system}.

First, we have a 1\,000-actuators MEMS DM dedicated to the correction of NCPAs, called non-common path corrector (NCPC) DM. Such errors are invisible to the main AO loop, as they are not in the same optical path as the PWFS. The NCPC DM can therefore be used to correct the errors introduced in the coronagraphic arm, as well as the residuals errors not corrected by the primary DMs. After the NCPC DM, a filter wheel contains the aperture pupil stops, including the `bump mask'. Subsequently, the beam passes through a specially designed mount containing the first set of aspheric PIAA lenses. A PIAA set contains two aspheres. Each asphere is mounted in its own X-Y stage with motorised actuators allowing for precise alignment control. This is fixed to a linear stage that can move the mount in and out of the beam. The individual mounts and the linear stage together give five degrees of freedom to align the PIAA lenses. An off-axis parabola brings the beam into a focus onto a coronagraphic focal plane mask. Masks are placed in a filter wheel after the PIAA stage. MagAO-X currently supports classic Lyot coronagraphs with opaque focal plane masks and PIAACMC. Any light reflected from the mask is sent to a separate focal plane low-order wavefront sensor (LOWFS, \citet{lowfs_guyon}). Another downstream filter wheel contains the Lyot stops. Reflective Lyot stops are used to send the rejected light to a dedicated Lyot LOWFS sensor \citep{lyot_lowfs}. An additional stage with the second set of PIAA lenses is mounted in the same way as the first set, and can move in or out of the beam. Finally, the two science cameras have a filter wheel with neutral density filters and a beam splitter to divide the light between them.

\begin{figure*}[h!]
    \centering
    \includegraphics[width=0.95\linewidth]{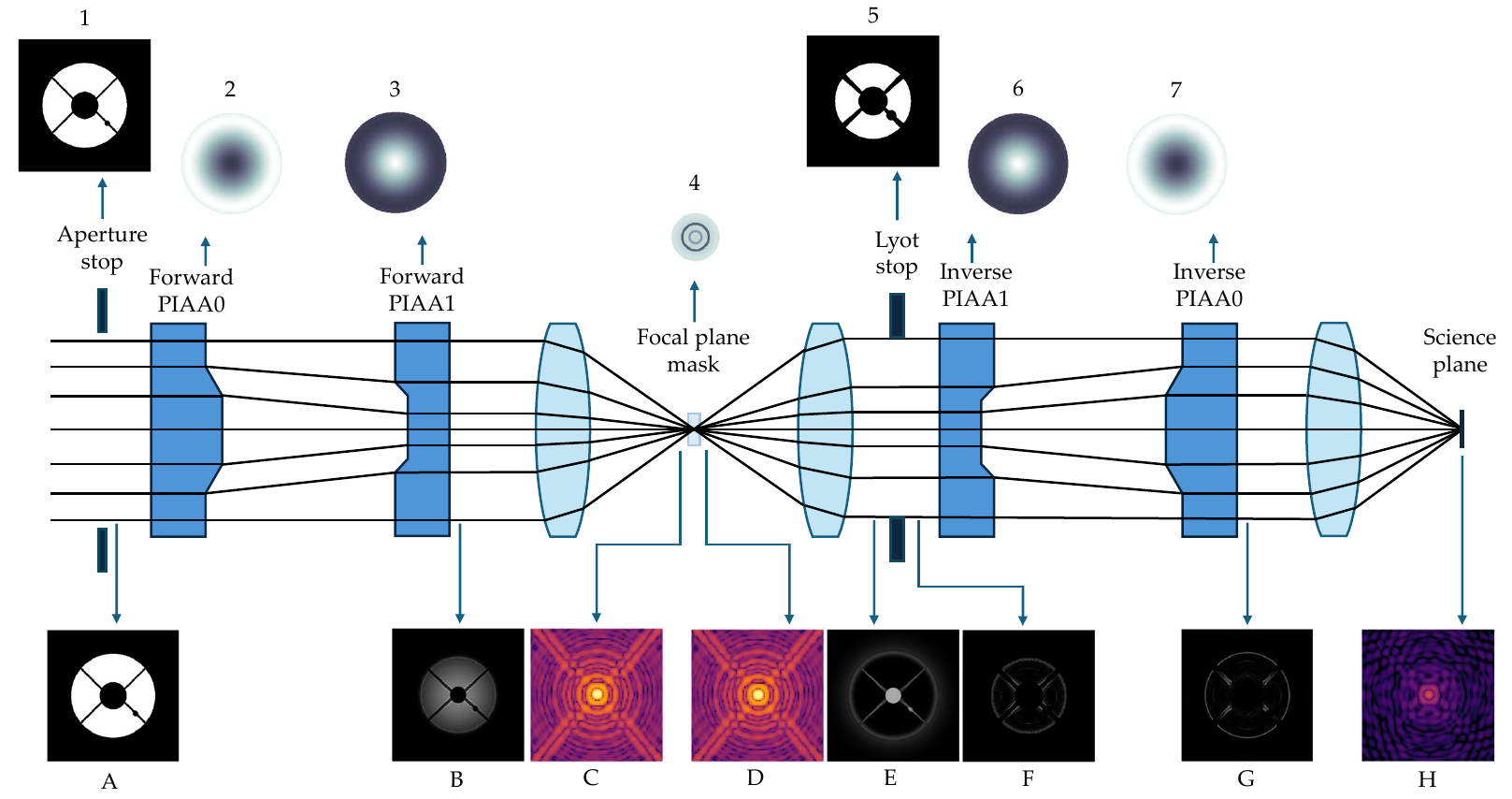}
    \caption{PIAACMC optical layout. The top figures show the simulated components designed for MagAO-X, and the bottom figures show the simulated pupil or focal plane images at different positions in the system, for an on-axis source. Top figures from left to right: (1) MagAO-X's binary aperture stop used for coronagraphic observations (`bump mask'); (2) forward PIAA0, the first aspheric PIAA lens of the forward set; (3) forward PIAA1, the second aspheric PIAA lens of the forward set; (4) phase-shifting focal plane mask, a transmissive optical component to suppress on-axis light; (5) MagAO-X's Lyot stop for PIAA observations, a binary mask to block the light diffracted by the focal plane mask, modelled on the apodised pupil shape; (6) inverse PIAA1, the first aspheric PIAA lens of the inverse set. It is the same as foward PIAA1 but flipped; (7) inverse PIAA0, the second aspheric PIAA lens of the inverse set. It is the same as forward PIAA0 but flipped. Bottom figures from left to right: (A) light distribution at the MagAO-X aperture; (B) light distribution after the forward PIAA lenses set; (C) the MagAO-X point spread function (PSF) on the focal plane; (D) PSF after the focal plane mask. It doesn't change in intensity because the focal plane mask is transmissive and only acts on the phase of the light; (E) light in the pupil plane that is diffracted by the focal plane mask; (F) residual light in the pupil plane after the Lyot stop; (G) residual light in the pupil plane after de-apodisation by the inverse PIAA lenses set; (H) residual light in the science plane.}
    \label{fig:PIAACMC_layout}
\end{figure*}

\subsection{PIAACMC layout}\label{subsec:PIAACMC}
The PIAACMC \citep{Guyon2010}) is a small IWA, high-throughput coronagraph that closely approaches the coronagraphic theoretical performance limit. This holds even when considering extended central sources \citep{Guyon2006} and entrance pupils with complex shapes \citep{Guyon2013}. Figure \ref{fig:PIAACMC_layout} shows the optical layout of the PIAACMC.

First, a set of two aspheric lenses (called the forward PIAA set, including the first lens, PIAA0, and the second lens, PIAA1) apodise the entrance pupil without loss in throughput. The apodisation is performed via geometrical remapping of the flux within the pupil plane. Contrarily to classical apodisation that blocks part of the light and degrades the angular resolution, PIAA optics maintain full throughput and resolution. The first lens mainly redistributes the light across the pupil by introducing phase differences to create the apodisation profile, while the second lens collimates the outgoing rays and corrects for such phase differences to return to a flat wavefront. However, the redistribution and correction tasks are partly shared between the two lenses. The optimal distribution profile for coronagraphic nulling is given by a prolate spheroidal function whose parameters depend on the desired size of the focal plane mask. An overview of the PIAA technique and the lens surface sag equations for flux remapping are detailed in \citet{Guyon2003} and \citet{Guyon2005}. The process for designing PIAA lenses functions for complex apertures is detailed in \citet{Guyon2012}. The forward PIAA lenses create an unaberrated apodised wavefront only for an on-axis source. For off-axis sources, the light hits the second lens at a different angle and position, causing imperfect correction of the phase differences introduced by the first lens. Therefore, the focal image of an off-axis source is distorted, with a shifted Airy core and a faint `tail' pointing towards the optical axis \citep{Guyon2005}, creating a shape that resembles a pineapple. A set of inverse PIAA lenses is introduced after the Lyot stop to compensate for residual phase differences and regain diffraction-limited imaging capabilities. These lenses are exactly the same as the forward PIAA set but flipped and in inverse order (first the flipped PIAA1 lens and then the flipped PIAA0 lens).

The focal plane mask (FPM) acts on the complex electric field and, particularly, on the phase of the light. It does not absorb or reflect on-axis light, but rather phase-shifts it to create destructive interference. Off-axis light misses the mask and propagates unperturbed through the system. The optimal size of the FPM is dependent on the strength of the apodisation that is created by the PIAA lenses. However, the mask's radius must be larger than about 0.54 $\lambda$/D, a critical value that enables full on-axis light extinction \citep{Guyon2010}.

Finally, a Lyot stop \citep{lyotstop} must be placed in the pupil plane before the inverse PIAA set. The Lyot stop is needed to block the light that was diffracted outside of the geometrical pupil by the FPM. The shape of the Lyot stop should therefore match the shape of the entrance pupil of the system after the apodisation and should be slightly oversized. This enables the effective blocking of on-axis diffracted light while as much off-axis light as possible propagates to the science plane.

Thanks to the lossless apodisation and the use of a phase mask, the PIAACMC enables imaging at the diffraction limit, with a potential sub-$\lambda$/D IWA and high throughput. However, this concept also comes with a few challenges, which are currently the manufacturing precision of the FPM and its robustness to tip-tilt errors.

\section{Design and manufacturing}\label{sec:design_manufacturing}
\subsection{Design}\label{subsec:design}
\subsubsection{PIAA lenses and Lyot stop}
MagAO-X's PIAA lenses and Lyot stop design and implementation was performed by \citet{warren_thesis}. The design process is iterative and consists of propagating the light multiple times through the elements of the system while updating their properties until the optimal design is reached. This was performed using the software Coronagraph Optimization For Fast Exoplanet Exploration (COFFEE\footnote{\url{https://github.com/coffee-org/coffee}}) developed by the SCExAO group.

The PIAA lenses can be represented as a composition of different modes describing their surfaces. The optimisation process only tunes a certain number of these modes at each iteration. The light is propagated through a simple FPM with a size determined uniquely by the apodisation parameters given as input to the PIAA lenses. Finally, the Lyot stop is numerically optimised by searching for different locations and shapes that provide extinction of the residual diffracted light.

\begin{figure}[h!]
    \centering
    \includegraphics[width=0.95\linewidth]{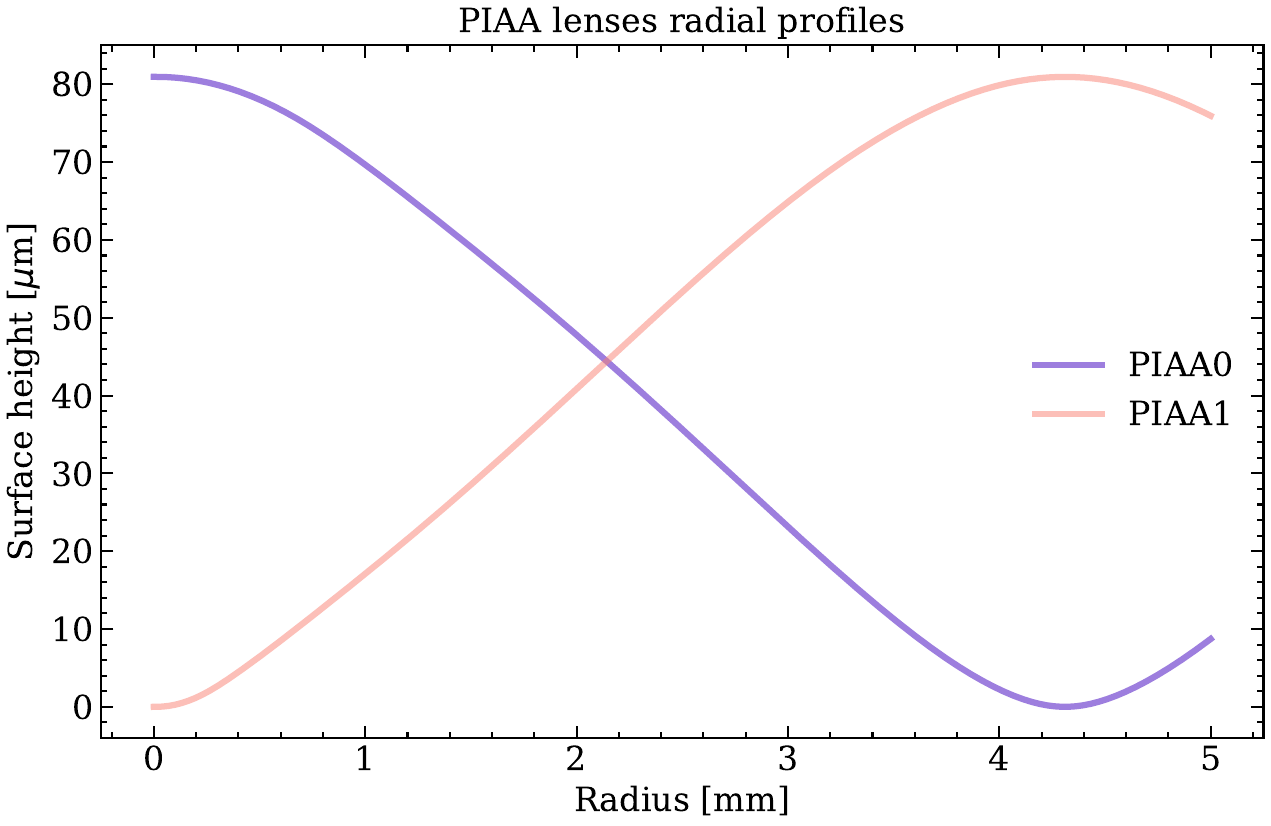}
    \caption{Surface heights of the PIAA lenses designed for MagAO-X by \citet{warren_thesis}. The dark purple line (PIAA0) is the height of the first lens of the forward PIAA set, and the light pink line (PIAA1) is the height of the second lens of the forward PIAA set. The inverse PIAA lenses set has the same surface heights with the lenses inverted.}
    \label{fig:PIAA_surface_curves}
\end{figure}

The input values given to the PIAA lens optimiser are the system parameters (including geometry, $f$ number, source angular size, etc. of MagAO-X), the specifics for the lenses, including central wavelength and bandwidth ($\lambda$=900 nm $\pm$1\%), material (calcium fluoride, CaF\textsubscript{2}), and lens spacing (60 mm between the two surfaces). The surface profiles of the optimised PIAA lenses for MagAO-X are shown in Fig. \ref{fig:PIAA_surface_curves}.

\subsubsection{Focal plane masks}
\begin{figure*}[h!]
    \centering
    \includegraphics[width=0.95\linewidth]{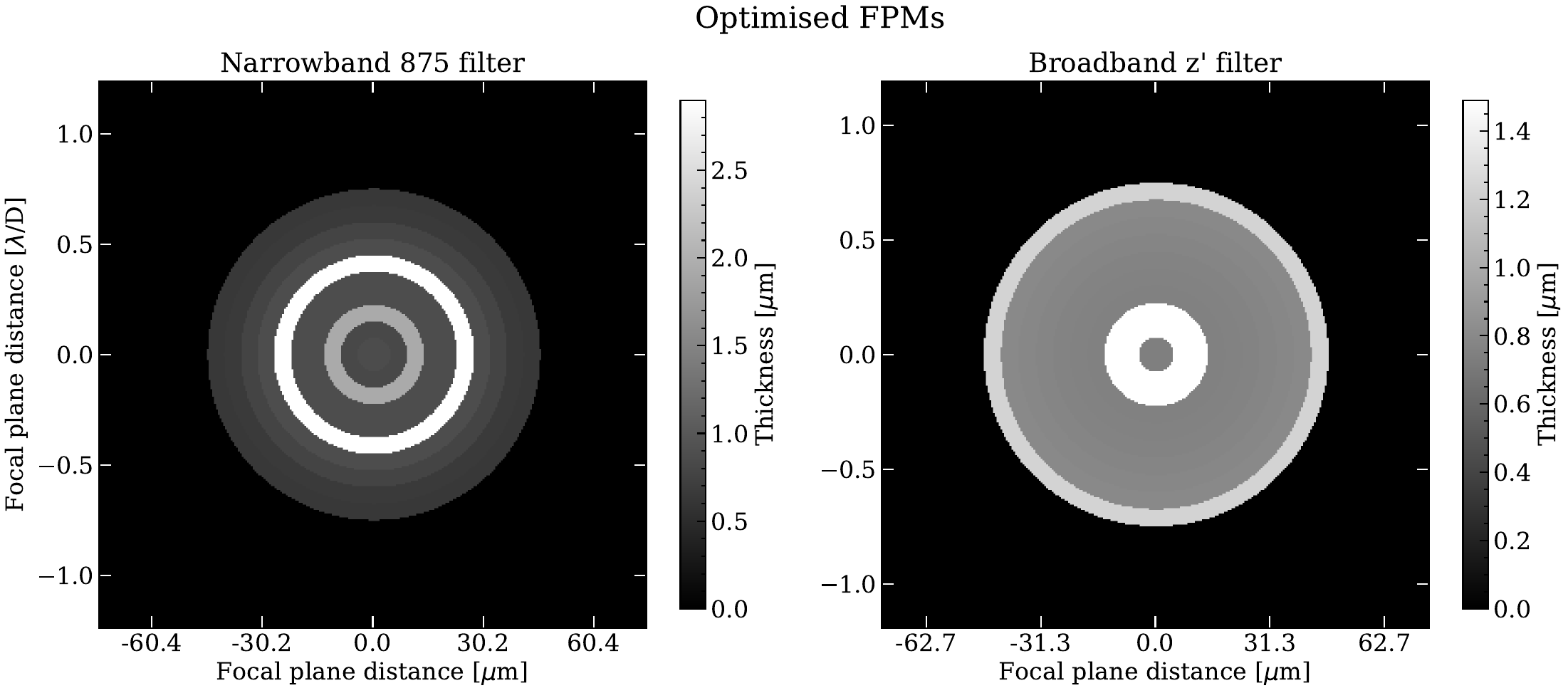}
    \caption{Optimised FPMs with ten concentric rings, for the narrowband 875 filter on the left, and for the broadband z' filter on the right. Note that the colourbars have different scales, and the X and Y axes are dependent on central wavelength of the filter. The axes are shown in both $\lambda$/D and physical units. Both masks have a radius of  0.75 $\lambda$/D, corresponding to 45.3 $\mu$m for the 875 filter and 47 $\mu$m for the z' filter.}
    \label{fig:FPM_designs}
\end{figure*}

The FPMs were designed by using a custom optimisation code developed in-house. The MagAO-X system was simulated entirely with the package High Contrast Imaging for Python, HCIPy\footnote{\url{https://docs.hcipy.org}} \citep{HCIPy}.
We designed masks for two MagAO-X filters, a narrowband filter at 875 nm $\pm$13 nm ($\sim$3\% bandwidth) and a broadband filter in z' at 908 nm $\pm$65 nm ($\sim$14\% bandwidth). In both cases, we set the radius of the mask to be 0.75 $\lambda$/D, where $\lambda$ is the central wavelength of the filter and D is the telescope diameter at the MagAO-X entrance pupil, while the $f$ number of MagAO-X is $f/69$. This corresponds to a physical FPM radius of 45.3 $\mu$m and 47 $\mu$m for the narrowband and broadband cases, respectively. These sizes were found to be optimal through simulations with FPMs of different sizes by evaluating their performance as described in the next paragraph.
The mask was parameterised into a number of N concentric rings where each ring has a different physical height. To keep the computation times reasonable, we clamped the maximum number of rings in the FPM to ten, since our simulations revealed no significant performance increases beyond this amount. We simulated the mask as a surface apodiser made of the resin IP-Dip with the refractive index defined in \citet{IPDIP_n}.

The optimisation process of the thicknesses of the different rings was done through minimisation of a cost function defined as the sum of the electric field power of an on-axis source in the science image.
To find the minimum of such function we used SciPy\footnote{\url{https://scipy.org/}} \citep{SciPy} and specifically the Newton-CG method that finds the global minimum through a basin hopping algorithm. The gradients were calculated analytically instead of numerically to speed up the optimisation process. We implemented reverse-mode algorithmic differentiation using the relations detailed in \cite{autodiff}.
These gradients are solely dependent on the relations between elements of the system. This means that they only need to be computed once at the beginning of the optimisation process. Since this optical system is linear, every operation can be represented by a matrix product. So, in the simulation, the wavefront is back-propagated through the optical elements to trace back the gradients.

The designs for the narrow and broadband cases are shown in Fig. \ref{fig:FPM_designs}. The full region shown in the figure is 150$\times$150 $\mu$m large, and the X-Y pixel resolution is 500 nm, so the figures are 300$\times$300 pixels in size.

\subsection{Manufacturing}\label{subsec:manufacturing}
\subsubsection{PIAA lenses}
The PIAA lenses were manufactured by Optimax. They were coated with a broadband anti-reflective coating with an average reflectance R\textsubscript{avg}=0.18\% between 600 nm and 1\,000 nm. The tolerance on surface irregularities was a root means square (RMS) of at most 20 nm, and the measured results for the four lenses is between 14 nm and 18.4 nm. Currently, we see no indication of the manufacturing errors of the lenses in the contrast results, and the residual NCPAs found are of the same order of those of other coronagraphs.

\begin{figure*}[h!]
    \centering
    \includegraphics[width=0.95\linewidth]{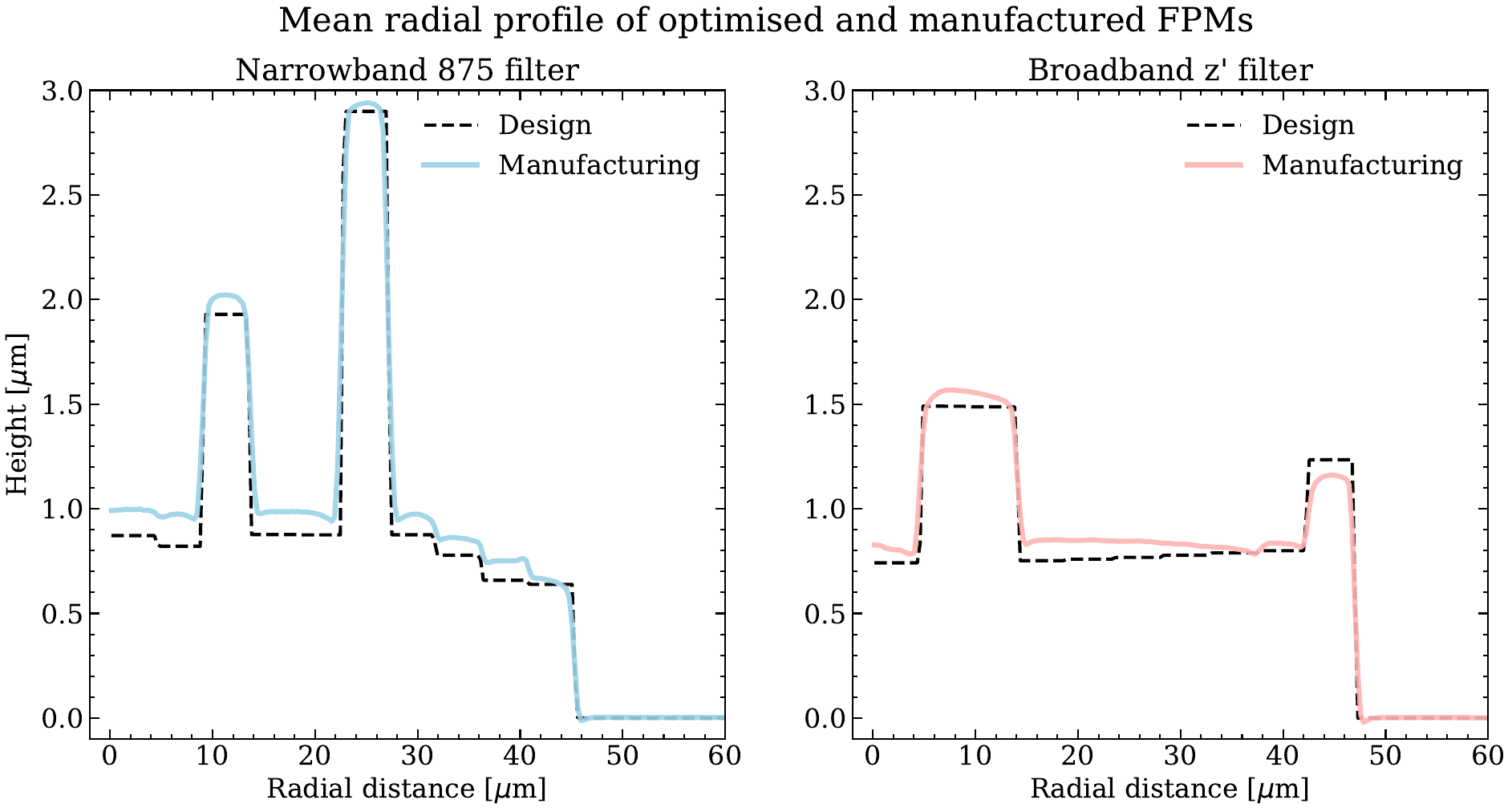}
    \caption{Mean radial profile of optimised and manufactured FPMs, for the narrowband 875 filter on the left, and for the broadband z' filter on the right. In both panels, the dashed black line (design) shows the mask design's radial profile, and the solid coloured line (manufacturing) shows the measured radial profile of the manufactured mask.}
    \label{fig:FPM_measurements}
\end{figure*}

\subsubsection{Focal plane masks}
We manufactured the FPMs at the Kraft Lab within the Physics Department of Leiden University using the Nanoscribe Photonic Professional GT machine. This machine is a 3D printer that allows to perform additive manufacturing based on two-photon polymerisation (2PP) for microfabrication. With 2PP, polymerisation of the material is triggered only at the focal point of a near infrared laser where the density of photons is high enough to have two-photon absorption. This allows for precise manufacturing with sub-$\mu$m precision in all directions. We used the 63x immersion objective, which provides highest resolution, and the photo-resistant resin IP-Dip. The prints were made on a 1 mm thick ultraviolet fused silica uncoated substrate.

Before printing, the tilt angle of the substrate was measured to compensate for writing errors. A manual realignment step was added if the measured angle was larger than 0.1\degr, because it would cause errors in the optical path difference (OPD) that are too big for our requirements. The machine has an automatic interface finder function that sets a 0 reference point in the Z direction for precise positioning of the structures on the substrate. When the printing starts, the voxels are partly embedded in the substrate or would otherwise not adhere to it and float away. To reach the exact desired height of the structure while making it adhere to the substrate, an offset from this reference can be set manually. This is called Z offset. It is possible to find an optimal Z offset value that will minimise the portion of embedded voxel while it still sticks to the substrate. The value can be found through a calibration process that normally does not vary from one substrate to the other. However, due to an unknown issue with the machine, possibly an error during the interface finder function, it was not possible to define a repeatable optimal Z offset value that must be found for every substrate. This issue made it difficult to reach the exact desired heights for our prints. To mitigate this problem as much as possible, the same mask was printed several times on the same substrate but with different Z offset values. Doing this is fair as long as the masks are separated enough to prevent interference between them during observations. The mask that got closer to the design was picked for testing and observations.

We measured the masks' height profiles with a 3D laser scanning microscope at the Space Research Organisation Netherlands (SRON) and chose the best manufactured designs. Fig. \ref{fig:FPM_measurements} shows the mean radial profiles of the best manufactured masks. Their root mean square error (RMSE) in Z direction is $\sim$60 nm and $\sim$40 nm, for the narrowband and broadband case, respectively. This corresponds to a surface deviation of the order of $\sim$100 nm. We expect this to deteriorate the performance, making the on-axis null less efficient, and thus worsening the reachable contrast.

We performed a tolerance analysis by simulating the masks' performance when random errors of different orders of magnitude are added. The errors impact Z offset, tilt angle, and ring heights, acting at the same time, and were simulated with a standard normal distribution. We ran different simulations and took their median to evaluate the performance degradation as a function of the errors. We estimate that the current designs would require an RMSE of the order of $\sim$5 nm to approach the desired performance, which translates to surface deviations in Z direction of the order of $\sim$10 nm. This assumes a substrate tilt error smaller than 0.1\degr and a very small error in Z offset (within 5 nm), both of which are feasible to reach in normal conditions. The lateral spatial resolution has a lower impact on the performance, and is well within our requirement. Currently it was not possible to reach the required tolerances in manufacturing mainly because of the machine malfunction we experienced.

Although this tolerance seems pretty small, we are confident that by using the machine at its best capabilities we can approach surface deviations of the order of $\sim$10 nm. From our simulations, we currently see that by correcting for the Z offset error, the mean surface deviation in height reduces to $\sim$50 nm. This writing error can possibly be improved even further by better calibrating the laser power, or might be avoided in the first place by improving the robustness of the mask's design and its sensitivity to writing errors.

\begin{figure*}[h!]
    \centering
    \includegraphics[width=0.95\linewidth]{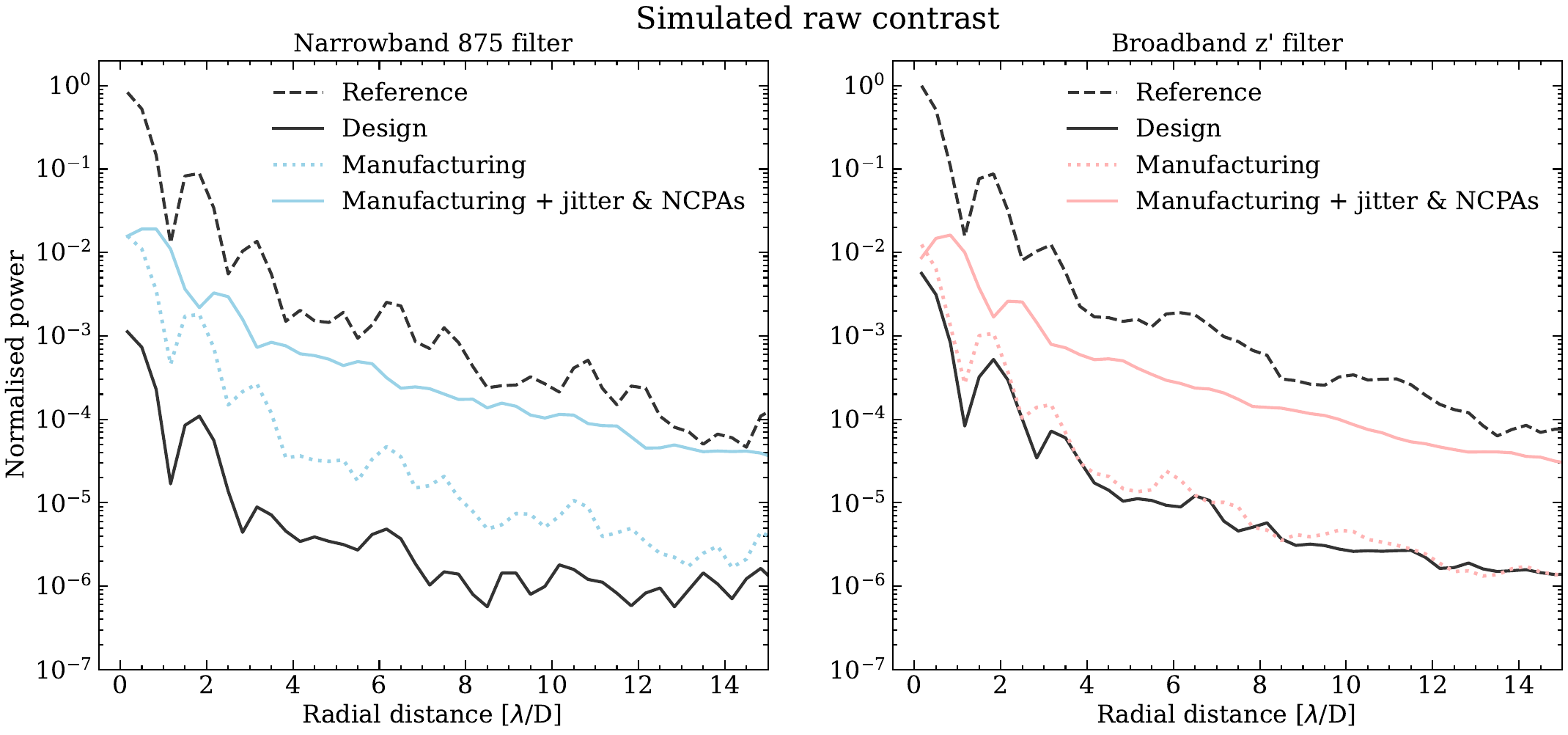}
    \caption{Simulated raw contrast curves as a mean radial profile, for the narrowband 875 filter on the left and the broadband z' filter on the right. In both panels, the solid black line (design) shows the simulated contrast with the optimised mask, the dotted coloured line (manufacturing) shows the simulated contrast with measurements of the manufactured mask, the solid coloured line (manufacturing + jitter \& NCPAs) shows the simulated contrast with measurements of the manufactured mask and perturbations including a jitter of 7 mas RMS and typical residual NCPAs from MagAO-X, and the dashed black line (reference) shows the simulated non-coronagraphic PSF. In this experimental setup, all elements are in the beam (bump mask, forward PIAA lenses set, focal plane mask, Lyot stop, inverse PIAA lenses set), and the focal plane mask is taken out of the beam only to measure the reference.}
    \label{fig:expected_contrast}
\end{figure*}

We implemented the printed masks in our simulations using their height measurements. For each filter, we selected which printed mask gives the lower raw contrast, and we used them for our experiments. Figure \ref{fig:expected_contrast} shows the simulated raw contrast reached with the optimal designed masks\footnote{The masks used in this work were optimised with a size mismatch between simulation and actual MagAO-X instrument. With the corrected value, we are currently able to reach a contrast performance of about one order of magnitude better for the narrowband filter than the one showed in the figure.} and with the best printed masks, both in perfect conditions (no aberrations, other optical components are perfect). Additionally, we simulated the effect of typical MagAO-X residual jitter and NCPAs on the performance of the PIAACMC when using the printed mask. We do this because PIAACMC masks are highly sensitive to tip-tilt errors and currently the MagAO-X performance limit is given by a residual jitter of 7.2 mas RMS and residual NCPAs at the level of $\sim$2$\times$10\textsuperscript{$-$3}. These are currently the major limiting factors in the contrast performance, and vibration control for MagAO-X is currently one of the high priority improvements for the system. In these simulation setups, all elements were used (bump mask, forward PIAA lenses set, focal plane mask, Lyot stop, inverse PIAA lenses set), and the focal plane mask was not included only to measure the reference. The performance of the manufactured components was estimated by using their measured height maps from the laser scanning microscope in the end-to-end physical optics model.

As expected from the manufacturing errors, the new simulated contrast is worse than the design for both filters, and especially in narrowband. In the narrowband case, the average contrast within 1 and 5 $\lambda$/D with the optimal designed mask is about 1.8$\times$10\textsuperscript{$-$5}. When we model the manufactured component in the simulation, we get a contrast of about 3$\times$10\textsuperscript{$-$4} instead. In the broadband case, the average contrasts within 1 and 5 $\lambda$/D are more similar to each other. With the optimal designed mask the contrast is about 9.1$\times$10\textsuperscript{$-$5} and with the manufactured mask it is  1.8$\times$10\textsuperscript{$-$4}. The performance we expect to measure in laboratory experiments is the raw contrast shown by the solid coloured curves, where we simulated the printed masks and add the effects of typical residual jitter and NCPAs from MagAO-X. In this case, we reach an average contrast between 1 and 5 $\lambda$/D of about 1.6$\times$10\textsuperscript{$-$3} in the narrowband case and 1.4$\times$10\textsuperscript{$-$3} in the broadband case.

The FPM design we optimised for the narrowband filter is almost one order of magnitude better at small separations than the broadband filter design. This is due to the inherent chromaticity problem of a phase shifting mask. In fact, its profile will be optimal only for a specific wavelength and the nulling capability will be worse for the other wavelengths in the band. So, the narrower the band, the better performance a phase-shifting mask can provide. The chromaticity problem cannot be completely removed but can be eased by relaxing the constraint in the optimisation code. Currently, the simulated contrasts with the masks that were manufactured are of the same order of magnitude for the two filters, which signals that the performance is currently mainly limited by the manufacturing precision. However, the design process must be improved to reach better contrasts. Some alternatives for improving the chromaticity issue will also be explored in future works.

\subsubsection{Lyot stop}
The Lyot stop was manufactured at the University of Arizona Nano Fabrication Center. This Lyot stop needs to be reflective to enable LOWFS \citep{lyot_lowfs}. The manufacturing process involved depositing a patterned gold coating on top of a fused silica substrate. The gold coating was only applied to the areas that need to be blocked by the Lyot stop.

\section{Laboratory performance characterisation}\label{sec:lab_results}
In this section, we show measurements made with MagAO-X by using its internal source, hence in `lab mode'. We analyse the performance at large off-axis angles and calculate throughput and contrast curves to characterise the performance of the PIAACMCs we manufactured.

\begin{figure*}[h!]
    \centering
    \includegraphics[width=0.95\linewidth]{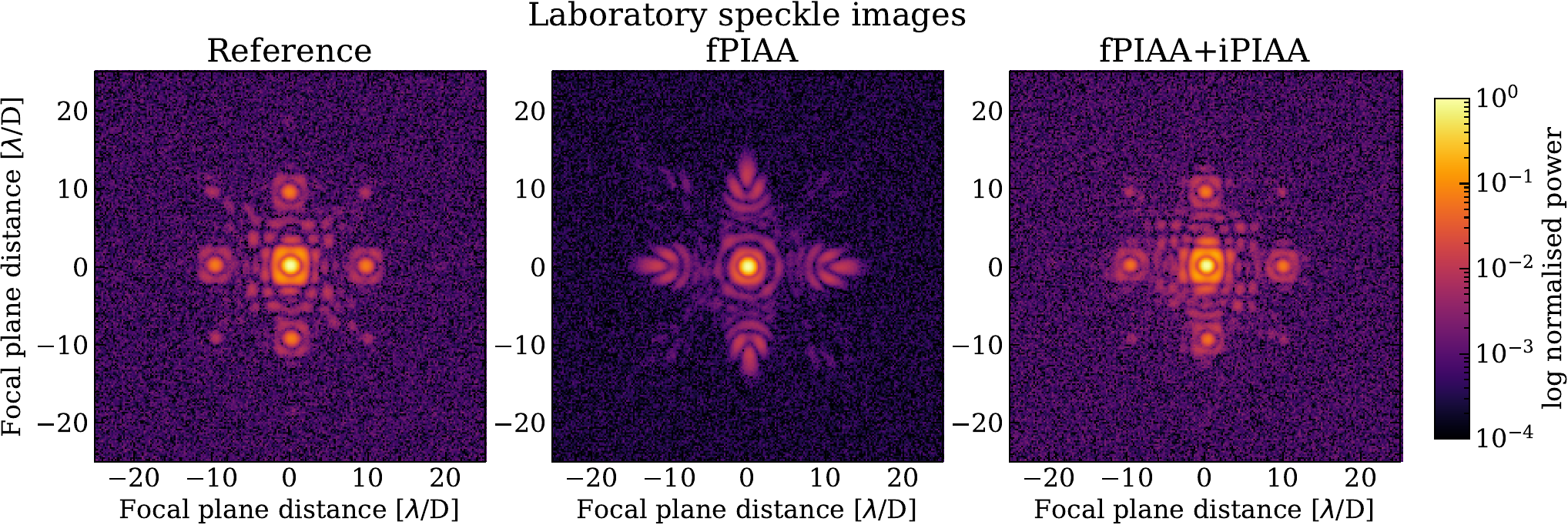}
    \caption{Examples of focal plane images of MagAO-X's internal source with artificial DM speckles at 10 $\lambda$/D separation for the following: The reference (left panel; only bump mask and Lyot stop in the beam), fPIAA (central panel; bump mask, Lyot stop, and forward PIAA lenses set in the beam), and fPIAA+iPIAA (right panel; bump mask, Lyot stop, and both forward and inverse PIAA lenses sets in the beam). The fPIAA image shows the typical off-axis distortions caused by the PIAA lenses. With the inverse PIAA lenses set the round shape of the off-axis speckles is restored.}
    \label{fig:pineapples}
\end{figure*}

\begin{figure*}[h!]
    \centering
    \includegraphics[width=0.95\linewidth]{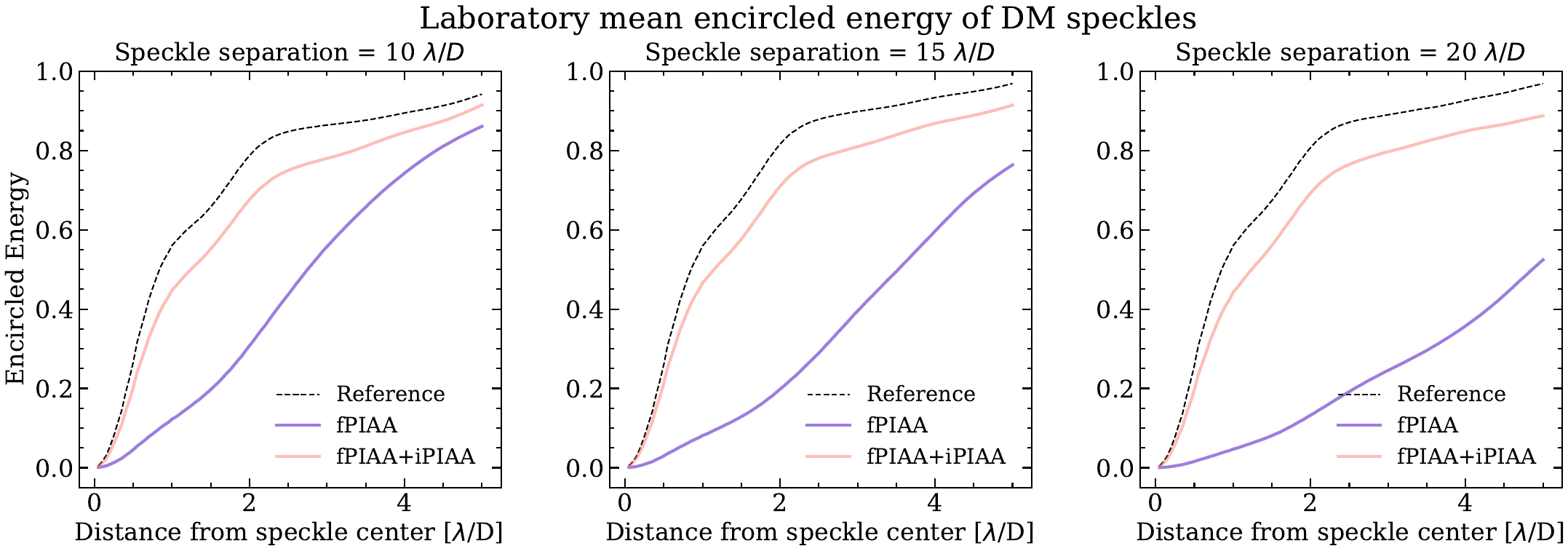}
    \caption{Laboratory mean encircled energy measurements of the four DM speckles at different separations: 10 $\lambda$/D (left panel), 15 $\lambda$/D (central panel), and 20 $\lambda$/D (right panel). In all panels, the dashed black line (Reference) shows measurements with only bump mask and Lyot stop in the beam; the dark purple line (fPIAA) shows measurements with bump mask, Lyot stop, and the forward PIAA lenses set in the beam; and the light pink line (fPIAA+iPIAA) shows measurements with bump mask, Lyot stop, and both forward and inverse PIAA lenses sets in the beam. The distortions caused by the forward PIAA lenses get worse for larger separations, as we deviate more from the optimal on-axis position. The inverse PIAA lenses restore the original encircled energy within a small margin between $\sim$97\% to $\sim$92\% depending on the DM speckle separation.}
    \label{fig:encircled_energy}
\end{figure*}

\begin{figure*}[h!]
    \centering
    \includegraphics[width=0.95\linewidth]{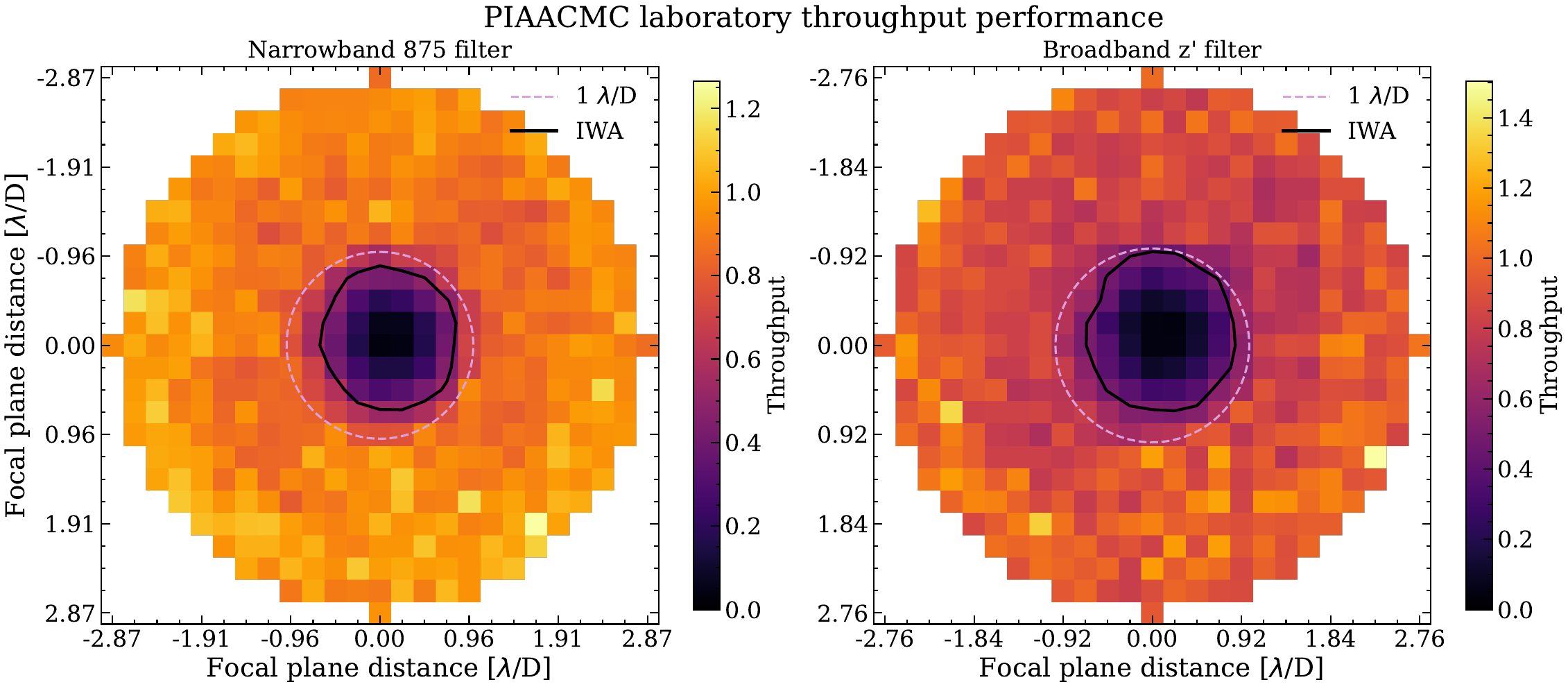}
    \caption{PIAACMC 2D laboratory throughput maps and their IWAs, for the narrowband 875 filter on the left, and the broadband z' filter on the right. In this experimental setup, all elements are in the beam (bump mask, forward PIAA lenses set, focal plane mask, Lyot stop, inverse PIAA lenses set). The dashed purple line shows 1 $\lambda$/D separations around the optical axis, and the solid black line shows where throughput is 0.5 (IWA). Note that, although the data was normalised for laser power variations, residual fluctuations are causing the computed throughput to exceed unity at a few scanned positions. The colourbars of the two panels are different.}
    \label{fig:TP_lab_2D}
\end{figure*}

\begin{figure*}[h!]
    \centering
    \includegraphics[width=0.95\linewidth]{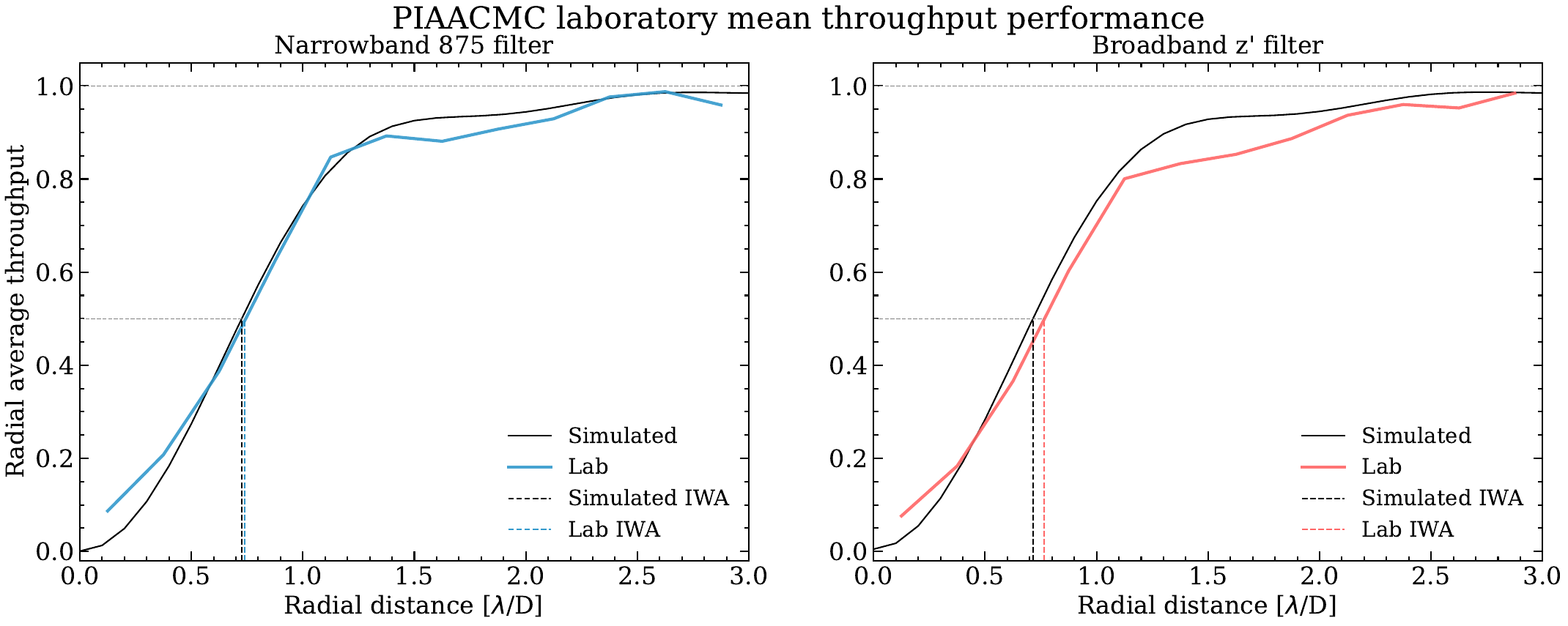}
    \caption{PIAACMC laboratory mean radial throughput curves and their IWAs, for the narrowband 875 filter on the left, and the broadband z' filter on the right. In this experimental setup, all elements are in the beam (bump mask, forward PIAA lenses set, focal plane mask, Lyot stop, inverse PIAA lenses set). In both panels, the solid black line (simulated) shows the simulated throughput with the measurements of the manufactured mask; the solid coloured line (lab) shows the laboratory throughput measurements with the internal source; and the dashed lines help identifying the IWA values in the simulated case (black) and the lab case (coloured).}
    \label{fig:TP_lab_curve}
\end{figure*}

\subsection{Off-axis performance}

We emulated off-axis sources at desired separations to study the performance of PIAACMC at off-axis angles. So, we created incoherent DM speckles by applying a modulated sinusoidal pattern on the tweeter DM \citep{Jovanovic2015}. The pattern applied to the DM is two sinusoids, along X and Y directions, creating four speckles in the focal plane. These all have the same amplitude and separation from the on-axis point, but are placed at 90\degr angles from each other. Their orientation can be controlled through the DM pattern.

We recorded data using three different configurations: (1) using only the bump mask and Lyot stop, called `Reference' (2) using bump mask, Lyot stop, and the forward PIAA lenses set, called `fPIAA', and (3) using bump mask, Lyot stop, and both forward and inverse PIAA lenses sets in the beam, called `fPIAA+iPIAA'. We took these measurements for different DM speckle separations, while keeping their amplitude and angle constant. Figure \ref{fig:pineapples} shows examples of these three measurements with DM speckles placed at a separation of 10 $\lambda$/D. Then, we computed the encircled energy by first determining the center position of each speckle and then calculating the total power within circles of increasing radius centred at those points. We normalised by an estimate of the total energy of the reference speckle, and show the mean of the four speckles in the encircled energy curves in Fig. \ref{fig:encircled_energy}. This figure shows the distortion of the off-axis sources as a function of distance from the central axis. Without inverse PIAA lenses, the encircled energy curve increases very slowly. We demonstrate that the performance can be restored with the inverse PIAA lenses for a wide range of separations. The encircled energy is only restored up to between about 91\% and 54\% without the inverse PIAA lenses, while with inverse PIAA lenses the encircled energy is restored up to between about 97\% and 92\%, for the three cases shown.

\subsection{Throughput and IWA}

\begin{figure*}[h!]
    \centering
    \includegraphics[width=0.95\linewidth]{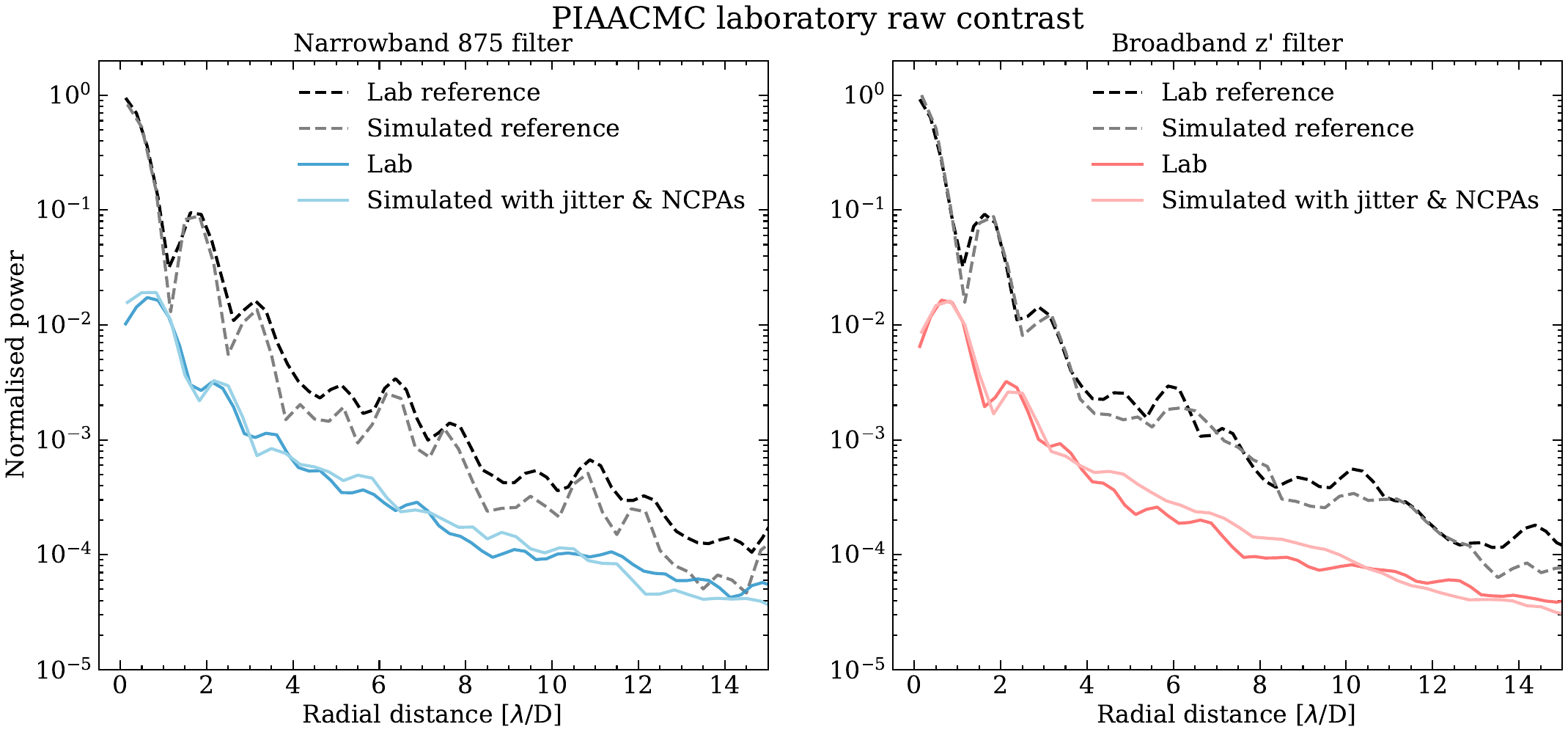}
    \caption{Raw contrast curves as a mean radial profile, measured with the instrument's internal source and simulated with the measurements of the manufactured mask, for the narrowband 875 filter on the left and the broadband z' filter on the right. In both panels, the grey dashed line (simulated reference) shows the simulated non-coronagraphic PSF, the lighter solid line (simulated with jitter \& NCPAs) shows the simulated contrast with measurements of the manufactured mask and perturbations including a jitter of 7 mas RMS and typical residual NCPAs from MagAO-X, the darker solid line (lab) shows the measured contrast with the internal source, and the dashed black line (lab reference) shows the measured non-coronagraphic PSF with the internal source. In this experimental setup, all elements are in the beam (bump mask, forward PIAA lenses set, focal plane mask, Lyot stop, inverse PIAA lenses set), and the focal plane mask is taken out of the beam only to measure the reference.}
    \label{fig:contrast_lab}
\end{figure*}

To create throughput curves, we first aligned the coronagraph with the internal source in the on-axis position, then we scanned the source in X and Y directions on the detector. This allows us to estimate the nulling capabilities of the masks for sources at different separations. In this experimental setup, all elements were in the beam (bump mask, forward PIAA lenses set, focal plane mask, Lyot stop, inverse PIAA lenses set). The scan was performed by applying various amounts of tip or tilt on the NCPC DM. We scanned from about $-$3  $\lambda$/D to 3 $\lambda$/D from the optical axis in both X and Y direction and discarded combinations that saturated the NCPC DM. We used DM spots to calibrate the power between images. These spots originate from periodic patterns on the DM (the regular distance between actuators), so they are always found in the same positions in the science image and their total power should be constant. The internal source power fluctuates, but the relative power of the spots stays the same, so we can use this to apply a frame-by-frame power normalisation. The throughput was computed as
\begin{equation}
    TP[x,y]=\frac{\sum \mathrm{obs}[x,y]}{\sum \mathrm{ref}} \frac{\mathrm{power_{ref}}}{\mathrm{power}[x,y]},
\end{equation}
where $TP[x,y]$ is the throughput at the coordinates $[x,y]$ from the on-axis central position, $\mathrm{obs}[x,y]$ is the PSF core image at those coordinates, and $\mathrm{power}[x,y]$ is the power factor computed from the DM spots for the image at those coordinates; $\mathrm{ref}$ is the reference PSF core, and $\mathrm{power_{ref}}$ is the power factor computed from the DM spots for the reference image. As the reference, we recorded unocculted data (taking the FPM out of the bean, everything else in the system stays the same) with the point source centred on axis, and computed an average image. All frames were bias subtracted for calibration purposes.

Figure \ref{fig:TP_lab_2D} shows the 2D throughput maps and highlights the IWAs. Here, because of some residual laser fluctuations, the throughput exceeds unity at a few scanning positions. Moreover, the IWA curves we find are slightly asymmetric, which might be due to an additional tip-tilt pattern that was applied on the NCPC DM. Figure \ref{fig:TP_lab_curve} does the same, but shows the throughput as a radial average. Both figures show that our PIAACMCs possess IWAs at the sub-$\lambda$/D level, specifically about 0.74 $\lambda$/D and 0.76 $\lambda$/D, respectively for the narrowband and broadband case. Since the results are generally very similar in the two cases, it seems that the bandwidth does not impact significantly the throughput performance of PIAACMCs. The measured results also follow the simulated throughput curves of the mask designs, which show IWAs of about 0.72 $\lambda$/D and 0.71 $\lambda$/D, respectively for the two filters.

\subsection{Contrast performance}
In this experimental setup, all elements were in the beam (bump mask, forward PIAA lenses set, focal plane mask, Lyot stop, inverse PIAA lenses set), and the focal plane mask was taken out of the beam only to measure the reference.

The contrast curves were created by computing the radial average power with the coronagraph in the beam. Their power normalisation was performed by using the incoherent DM speckles. This is possible by keeping their amplitude and position constant throughout all observations. The contrast curves are shown in Fig. \ref{fig:contrast_lab}. We computed the normalised raw contrast curve as
\begin{equation}
    C_{\mathrm{norm}}=\frac{\mathrm{obs}}{\mathrm{power_{obs}}}\frac{\mathrm{power_{ref}}}{\mathrm{max[ref]}},
\end{equation}
where $\mathrm{obs}$ is the scientific observation data with PIAACMC and $\mathrm{power_{obs}}$ is its power normalisation factor computed with the DM speckles; $\mathrm{ref}$ is the reference image, and $\mathrm{power_{ref}}$ is its power normalisation factor. The normalised reference PSF is simply the non-coronagraphic PSF divided by its maximum value.

The measure an average raw contrast with the internal source within 1 and 5 $\lambda$/D of about 1.6$\times$10\textsuperscript{$-$3} with the narrowband filter. With the broadband filter and within the same separation range we find a contrast of about 1.3$\times$10\textsuperscript{$-$3}. We expected the performance to be similar in the two filters because, as we mentioned, the main performance limit at small separation is currently the FPM manufacturing precision. Moreover, there results agree extremely well with the performance we expected from mask manufacturing errors and residual jitter and NCPAs.

\begin{figure*}[h!]
    \centering
    \includegraphics[width=0.95\linewidth]{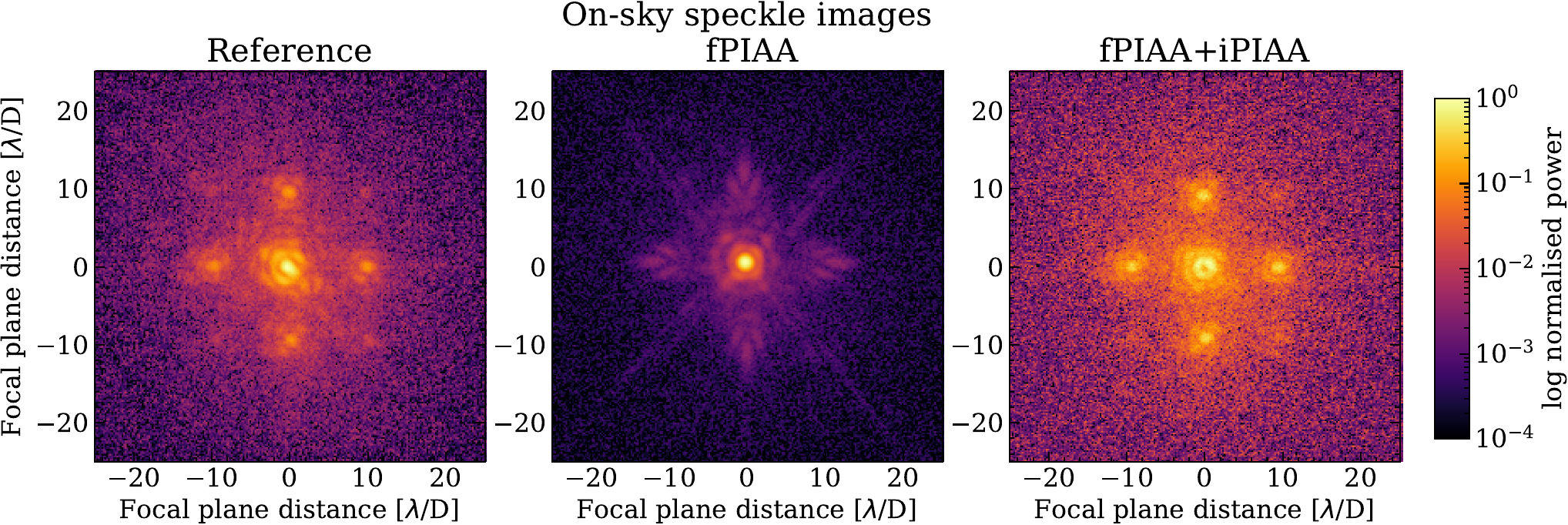}
    \caption{Examples of on-sky focal plane images with artificial DM speckles at 10 $\lambda$/D separation for the following: The reference (left panel; only bump mask and Lyot stop in the beam), fPIAA (central panel; bump mask, Lyot stop, and the forward PIAA lenses set in the beam), and fPIAA+iPIAA (right panel; bump mask, Lyot stop, and both forward and inverse PIAA lenses sets in the beam). Mind that in the right-most panel the focal plane mask is also in the beam. The inverse PIAA lenses restore the round shape of the off-axis speckles.}
    \label{fig:pineapples_onsky}
\end{figure*}

\begin{figure*}[h!]
    \centering
    \includegraphics[width=0.95\linewidth]{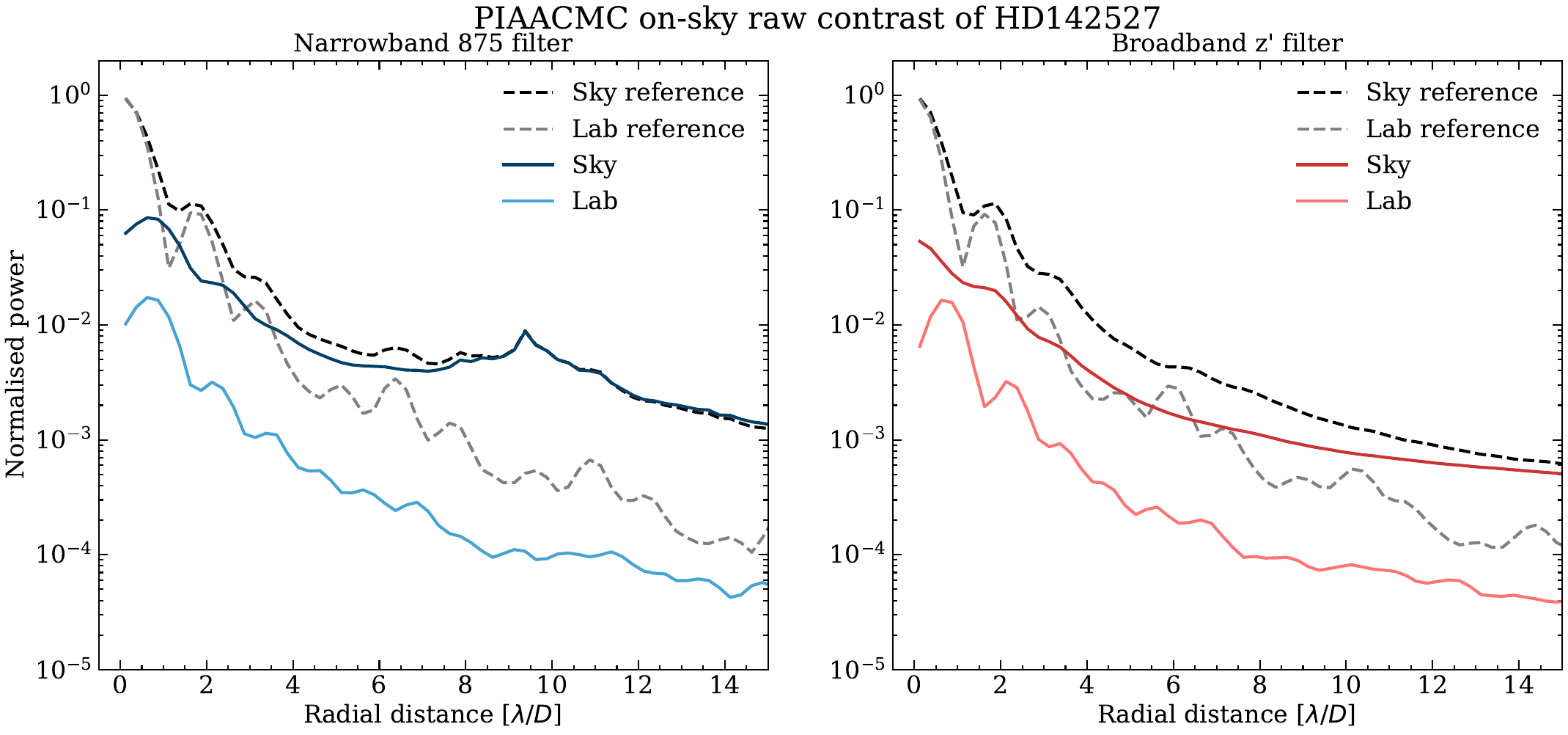}
    \caption{Raw contrast curves as a mean radial profile, for on-sky observations of HD\,142527 and measured with the instrument's internal source, for the narrowband 875 filter on the left and the broadband z' filter on the right. In both panels, the darker solid line (sky) shows the on-sky contrast, the lighter solid line (lab) shows the measured contrast with the internal source, the dashed black curve (sky reference) shows the on-sky non-coronagraphic PSF, and the dashed grey line (lab reference) shows the measured non-coronagraphic PSF with the internal source. In this experimental setup, all elements are in the beam (bump mask, forward PIAA lenses set, focal plane mask, Lyot stop, inverse PIAA lenses set), and the focal plane mask is taken out of the beam only to measure the reference. In the left panel, the peak at $\sim$9 $\lambda$/D in the on-sky curves is due to artificial DM speckles that were mistakenly not turned off.}
    \label{fig:contrast_onsky}
\end{figure*}

\section{On-sky performance characterisation}\label{sec:onsky_results}
Figure \ref{fig:pineapples_onsky} shows a PIAACMC alignment demonstration on-sky. We used the same alignment procedure used in lab mode, which seems to qualitatively look similar.

We observed the binary star HD\,142527 during the 2025A MagAO-X observing run, on the night of the 17\textsuperscript{th} of April 2025 starting at 08:27 UTC, with both filters. We closed the AO loop on 1\,563 modes, but with sub-optimal atmospheric conditions. Although the observatory's telemetry recorded a seeing measurement of $\sim$0.5 arcsec, we suspect that the conditions were further deteriorated by high wind speed in the upper atmosphere.

The on-sky contrast curves are shown in Fig. \ref{fig:contrast_onsky} and the lab contrast curves are also shown as a reference. In this experimental setup, all elements were in the beam (bump mask, forward PIAA lenses set, focal plane mask, Lyot stop, inverse PIAA lenses set), and the focal plane mask was taken out of the beam only to measure the reference. The peak at about 9 $\lambda$/D in the observations with the 875 filter is due to artificial DM speckles that were mistakenly not turned off. We reach an average raw contrast within 1 and 5 $\lambda$/D of about 1.4$\times$10\textsuperscript{$-$2} with the narrowband filter, and of about 7.8$\times$10\textsuperscript{$-$3} for the broadband filter. This means that we lose about one order of magnitude of on-axis light suppression on-sky for both filters with respect to the lab results. The likely reason the degraded performance is the imperfectly corrected atmospheric turbulence and atmospheric conditions. Again, we are mainly limited by quasi-static speckles in MagAO-X that are not properly corrected, especially at separations larger than about 2-3 $\lambda$/D. The lab and on-sky non-coronagraphic measurements (reference curves) in fact also differ from each other, the latter being smoother and reaching lesser contrast.

\section{Conclusions}\label{sec:conclusions}
We designed and manufactured two focal plane masks for the PIAACMC and tested them with MagAO-X. We successfully characterised the on-sky PIAACMC contrast and IWA performance for the first time. This was also the first time a PIAACMC was tested on-sky at sub-micron NIR wavelengths. We showed results with a narrowband filter (875 nm $\pm$13 nm) and with a broadband filter (908 nm $\pm$65 nm). This technological demonstration is an essential step in the development of small-IWA coronagraphs for the future generation of ELTs.

We described the design and manufacturing processes of PIAA lenses, FPMs, and Lyot stop for MagAO-X, and analysed the manufacturing errors of the phase shifting masks. We showed encircled energy plots with the internal source, demonstrating a good recovery of the original off-axis PSF shapes when both forward and inverse PIAA lenses set are aligned, within $\sim$92\% and $\sim$97\% depending on the separation. We also demonstrated sub-$\lambda$/D IWAs with internal source measurements of about 0.74 $\lambda$/D for the narrowband filter and 0.76 $\lambda$/D for the broadband filter. This result is particularly exciting in the context of a new era of extremely large ground-based observatories. The average raw contrast reached with the internal source of the instrument within 1 and 5 $\lambda$/D is about 1.6$\times$10\textsuperscript{$-$3} with the narrowband filter and 1.3$\times$10\textsuperscript{$-$3} with the broadband filter. These agree extremely well with the simulated contrasts when using the manufactured masks and adding the effects of residual jitter and NCPAs of MagAO-X. Residual jitter and NCPAs are currently the major limiting factors in the contrast performance, and vibration control for MagAO-X is currently one of the high priority improvements for the system. We finally showed on-sky average raw contrasts within 1 and 5 $\lambda$/D of about 1.4$\times$10\textsuperscript{$-$2} with the narrowband filter and 7.8$\times$10\textsuperscript{$-$3} with the broadband filter. This loss of about one order of magnitude with respect to the lab measurements is most likely caused by a mix of poor atmospheric conditions, imperfect wavefront correction from the AO system, and leakage through the coronagraph from tip-tilt errors.

The masks' manufacturing process is not optimal yet due to a technical issue we encountered. Manufacturing more accurate FPMs by tweaking machine writing parameters will be pursued in the future to reach better contrast levels. Given the ready availability of Nanoscribe, we plan to continue using this technique in the future, but we might also explore alternative manufacturing processes, for instance lithography or liquid crystals.

Improving the FPM optimisation code is essential to enhance the performance reached with the PIAACMC. First, introducing expected manufacturing errors in the optimisation process will improve robustness of the masks. An additional step would be to consider spatially resolved stars instead of only point sources. Moreover, introducing tolerance against low-order aberrations or misalignments will allow us to reach better on-sky contrasts. Future works will also include integrating focal plane wavefront control such as implicit electric field conjugation (iEFC) for NCPA correction with the PIAACMC. This will allow us to go beyond the contrast limitations of MagAO-X's quasi-static speckles.

We will perform more on-sky observations with MagAO-X to further explore the performance of PIAACMCs. As shown, the PIAACMC provides extremely small IWAs, which makes it among the best candidate coronagraphs for future ground-based and space-based observatories. Specifically, it will enable coronagraphy at the diffraction limit with the ELTs, which will be essential to unlock a whole new population of temperate exoplanets that closely orbit their host star.

\begin{acknowledgements}
The MagAO-X phase II project acknowledges generous support from the Heising-Simons Foundation. We are very grateful for support from the NSF MRI Award \#1625441 (MagAO-X). SYH acknowledges support from NWO Award 184.036.004. JL and SYH acknowledge support from NASA APRA award 80NSSC24K0288.
\end{acknowledgements}

\bibliographystyle{aa} 
\bibliography{bibliography.bib} 

\end{document}